\documentclass[graybox]{svmult}

\usepackage{mathptmx}       
\usepackage{helvet}         
\usepackage{courier}        
\usepackage{type1cm}        
\usepackage{makeidx}         
\usepackage{graphicx}        
\usepackage{multicol}        
\usepackage[bottom]{footmisc}
\usepackage{amsmath}

\makeindex             

\begin{document}

\title*{Theories for influencer identification in complex networks}
\author{Sen Pei, Flaviano Morone and Hern\'{a}n A. Makse \\In {\it Complex Spreading Phenomena in Social Systems}, edited by Sune Lehmann and Yong-Yeol Ahn (Springer Nature, 2018)}
\authorrunning{Sen Pei, Flaviano Morone and Hern\'{a}n A. Makse}
\institute{Sen Pei \at Department of Environmental Health Sciences, Mailman School of Public Health, Columbia University, New York, NY 10032, USA, \email{sp3449@cumc.columbia.edu}
\and Flaviano Morone \at Levich Institute and Physics Department, City College of New York, New York, NY 10031, USA, \email{flaviomorone@gmail.com}
\and Hern\'{a}n A. Makse \at Levich Institute and Physics Department, City College of New York, New York, NY 10031, USA, \email{hmakse@lev.ccny.cuny.edu}}
%
%
\maketitle

\abstract{ In social and biological systems, the structural heterogeneity of interaction networks gives rise to the emergence of a small set of influential nodes, or influencers, in a series of dynamical processes. Although much smaller than the entire network, these influencers were observed to be able to shape the collective dynamics of large populations in different contexts. As such, the successful identification of influencers should have profound implications in various real-world spreading dynamics such as viral marketing, epidemic outbreaks and cascading failure. In this chapter, we first summarize the centrality-based approach in finding single influencers in complex networks, and then discuss the more complicated problem of locating multiple influencers from a collective point of view. Progress rooted in collective influence theory, belief-propagation and computer science will be presented. Finally, we present some applications of influencer identification in diverse real-world systems, including online social platforms, scientific publication, brain networks and socioeconomic systems.}

\section{Introduction}
\label{sec:1}

In spreading processes of information, it is well known that certain individuals are more influential than others. In the field of information diffusion, it has been accepted that the ability of influencers to initiate a large-scale spreading is attributed to their privileged locations in the underlying social networks \cite{watts2007influentials,kitsak2010identification,pei2013spreading,min2016searching}. Due to the direct relevance of influencer identification in such phenomena as viral marketing \cite{leskovec2007dynamics}, innovation diffusion \cite{rogers2010diffusion}, behavior adoption \cite{centola2010spread} and epidemic spreading \cite{pastor2001epidemic}, the research on searching for influential spreaders in different settings is becoming increasingly important in recent years \cite{pei2013spreading}.

In the relative simple case of locating individual influencers, given the rich structural information encoded in nodes' location in the network, it is straightforward to measure the influence of a single node using centrality-based heuristics. Over the years, a growing number of predictors have been developed and routinely employed to rank single node's influence in spreading processes, among which the most widely used ones include number of connection \cite{albert2000error}, k-core \cite{seidman1983network}, betweenness centrality \cite{freeman1978centrality} and PageRank \cite{brin1998anatomy}, just to name a few. Beyond this non-interacting problem, a more challenging task is to identify a set of influencers to achieve maximal collective influence. Originally formulated in the context of viral marketing \cite{richardson2002mining}, collective influence maximization is in fact a core optimization problem in an array of important applications in various domains, ranging from cost-effective marketing in commercial promotion, optimal immunization in epidemic control, to strategic protection against targeted attacks on infrastructures. In addition to the topological complexity of network structure, collective influence maximization is further complicated by the entwined interactions between multiple spreaders, which renders the aforementioned centrality-based approaches invalid. As a result, it is required to treat the problem from a collective point of view to develop effective solutions \cite{morone2015influence}.

\section{Finding individual influencers}
\label{sec:2}

In reality, many spreading phenomena are typically initiated by a single spreader. For instance, an epidemic outbreak in a local area is usually caused by the first infected person. For such processes, ranking the spreading capability of individual spreaders is of great significance in both accelerating and confining the diffusion.

\subsection{Topological measures}
\label{sec:2.1}

Intuitively, the nodes with large numbers of connections should have more influence on their direct neighbors. The disproportionate effect of highly-connected nodes, or hubs, on dynamical processes has been revealed in the early works on the vulnerability of scale-free networks \cite{albert2000error,cohen2001breakdown}. The targeted attack on a very small number of high-degree nodes will rapidly collapse the giant component of networks with heavy-tailed degree distribution. Compared with other more complex centrality measures, the computational burden of degree is almost negligible. Due to this, the simple degree centrality has been playing an important role in influencer identification.  In implementation, the performance of high-degree ranking can be further enhanced by a simple adaptive calculation procedure, that is, recalculating the degree of remaining nodes after the removal of previously selected nodes.

\begin{figure}[t]\centering
\includegraphics[width=1\columnwidth]{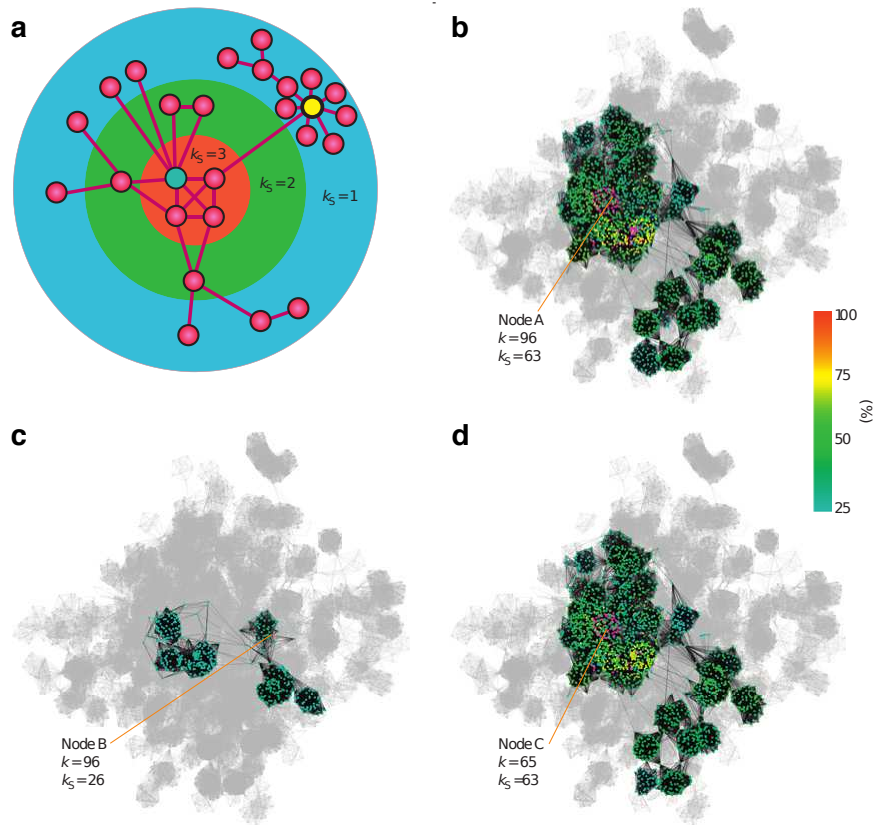}
\caption{{\bf a}, A schematic diagram of k-shell decomposition. The two highlighted nodes (blue and yellow), although both with degree $k=8$, are in different k-shells. {\bf b-d}, Infections starting from single nodes with same degree $k=96$ (A and B) can result in totally different outcomes. Whereas, infections originating from node C, locating in the same k-shell of node A ($k_S=63$) but with a smaller degree, are quite similar to the spreading from node A. The colors indicate nodes' probability to be infected in SIR simulations with infection rate $\beta=0.035$ and recovery rate $\mu=1$. Results are averaged over 10,000 realizations. Figure is adapted from Kitsak {\it et al.} \cite{kitsak2010identification}.}
\label{fig:1}
\end{figure}

An obvious drawback of degree centrality is that it only considers the number of direct neighbors. However, as indicated by empirical studies, most spreading phenomena are proceeded in a cascading fashion. Therefore, the ultimate influence of a single spreader is also affected by the global network structure. In realistic complex networks, high-degree nodes can appear at either the core area or the periphery region. This implies, the number of connections may not be a reliable indicator of influencers in real-world systems. Recently, Kitsak {\it et al.} confirmed this speculation through extensive simulations of susceptible-infected-recovered (SIR) and susceptible-infected-susceptible (SIS) dynamics on diverse real-world social networks \cite{kitsak2010identification}. In SIR model, a susceptible individual will become infected with a probability $\beta$ upon contact with his/her infected neighbors, and infected population will recover with a probability $\mu$ and become immune to the disease. In SIS model, the infection follows the same dynamics but infected persons will become susceptible again with a probability $\mu$. As shown in Fig. \ref{fig:1}b-d, SIR spreading processes initiated by two hubs with the same degree could result in quite different infected population, depending on their global position in the network. In contrast, the k-core index, which distinguishes the network core and periphery, is a more reliable predictor of influence. 

The k-core index is obtained by the k-shell decomposition in which nodes are iteratively pruned according to their remaining degree in the network (see Fig. \ref{fig:1}a) \cite{seidman1983network}. Specifically, nodes with degree $k=1$ are first removed successively until there is no node left with one link. The removed nodes are assigned with k-core index $k_S=1$. Then we remove nodes with degree $k=2$ similarly and continue to prune higher k-shells until no node left in the network. In terms of computational complexity, the above decomposition process can be finished within $O(M)$ operations, where $M$ is the number of links \cite{batagelj2003m}. Thus k-core ranking is feasible for large-scale complex networks encountered in big-data analysis. 

As illustrated in Fig. \ref{fig:1}a, the classification of k-core can be very different from that of degree. A hub with low k-core index is usually surrounded by many low-degree neighbors that limit the influence of the hub. On the contrary, nodes located in the core region, although may have moderate degree, are capable of generating large-scale spreading facilitated by their well-connected neighbors. In the case where recovered individuals do not develop immunity, infections would persist in the high k-core area. These findings challenge the previous predominate focus on the number of connections. The simple yet effective measure k-core has inspired several generalizations in consideration of the detailed local environment in the vicinity of high k-core nodes \cite{zeng2013ranking,liu2014core,liu2015improving,lu2016h}.

Although k-core was found effective in SIR and SIS spreading dynamics, some studies indicate that it may not be a good predictor of influence for other spreading models. For instance, in rumor spreading model, Borge-Holthoefer and Moreno \cite{borge2012absence} showed that the spreading capabilities of the nodes did not depend on their k-core values. These contradictory results relying on the choice of specific spreading model necessitate more extensive empirical validation with real information flow \cite{pei2014searching}.

Apart from the k-core index, another measure that takes into account the global network structure is eigenvector centrality \cite{bonacich1972factoring,restrepo2006characterizing}. The reasoning behind the eigenvector centrality is that the influence of an individual is determined by the spreading capability of his/her neighbors. Starting from a uniform score assigned to each node, the scores propagate along the links until a steady state is reached. In calculation, each step of score propagation corresponds to a left multiplication of the adjacency matrix to the current score vector. This procedure is actually the power method to compute the principal eigenvalue of the adjacency matrix. As a result, the steady score vector is in fact proportional to the right eigenvector corresponding to the largest eigenvalue. Notice that, supposing the initial score of each node is one, the first step of iteration will recover the degree centrality.

Despite the wide application of eigenvector centrality, it was recently found that the scores could be localized at a few high degree nodes due to the repeated reflection of scores from their neighbors during the iteration. Martin {\it et al.} solved this problem by using the leading eigenvector of the Hashimoto Non-Backtracking (NB) matrix \cite{martin2014localization}. In NB matrix, the immediate backtracking paths $i\to j$ and $j\to i$ are not permissible \cite{hashimoto1989zeta}, thus avoiding the heavy score accumulation caused by the recurrent one-step reflection. Recently, by mapping the SIR spreading process to bond percolation, Radicchi and Castellano proved that the NB centrality was an optimized predictor for single influencers in SIR model at criticality \cite{radicchi2016leveraging}. In next section, we will see the important role of NB matrix in collective influence maximization and optimal percolation \cite{morone2015influence}.

\subsection{Dynamics-based measures}
\label{sec:2.2}

Beyond the above pure topological measures, a number of centralities are developed on the basis of specific assumptions on the spreading dynamics. In some classical centralities proposed in the field of social networks, much emphasis is put on the shortest path. Along this way, several renowned centralities were developed and widely accepted in social network ranking. For instance, the closeness centrality quantifies the shortest distance from a given node to all other reachable nodes in the network \cite{sabidussi1966centrality}, while betweenness centrality measures the fraction of shortest paths cross through a certain individual between all node pairs \cite{freeman1978centrality}. A useful generalization of closeness centrality is the Katz centrality \cite{katz1953new}, which considers all possible paths in the network, but assigns a larger weight to shorter paths using a tunable parameter. In application, the applicability of these shortest-path-based centralities is limited by the high computational complexity of calculating the shortest paths between all pairs of nodes. As a result, they are more suitable for small or medium scale networks.

Another group of metrics are designed based on random walks. A famous random walk based centrality is PageRank \cite{brin1998anatomy}. As a revolutionary webpage ranking algorithm, PageRank mimics a random walk process along the directed hyperlinks. To avoid the random walker trapped in the dangled nodes, a jumping probability $\alpha$ is introduced to allow the walker jump to a randomly chosen node. The PageRank score is the stationary probability of each node to be visited by the random walker, which can be calculated through iteration. In applications, the PageRank of a node $i$ in a network can be calculated from $ p_t(i)=\frac{1-\alpha}{N}+\alpha\sum_{j}\frac{A_{ij}p_{t-1}(j)}{k_{out}(j)}$, where $k_{out}(j)$ is the number of outgoing links from node $j$ and $\alpha$ is the jumping probability. In a generalization called LeaderRank \cite{lu2011leaders}, a ground node is connected to all other nodes by additional bidirectional links. This procedure ensures the network to be strongly connected so that the convergence becomes faster.

In addition to the aforementioned centralities designed for general spreading processes, several measures are proposed aimed at specific dynamics, depending explicitly on model parameters. In these approaches, the development of measures is based on the equations depicting the dynamical process. Usually, the analysis of equations will naturally lead to the procedure of path counting in which the number of possible spreading paths is assessed. For instance, Klemm {\it et al.} developed a general framework to evaluate the dynamical importance (DI) of nodes in a series of dynamical processes \cite{klemm2012measure}. The iterative calculation of DI centrality essentially counts the total number of arbitrarily long walks departing from each node. Another metric relying on possible spreading paths is the expected force (ExF) proposed by Lawyer \cite{lawyer2015understanding}. To compute the expected force, all possible clusters of infected nodes after $n$ transmission events starting from a given node are enumerated. Then the entropy of their cluster degree (i.e., number of outgoing links of the cluster, or infected-susceptible edges) is calculated as the expected force for each node.

The approaches introduced here are far from complete. A growing number of metrics and methods are continuously proposed in the active area of finding single influencers \cite{lu2016vital}. In designing effective methods for more complex spreading models, the basic principles behind these measures should be universal.

\section{Finding multiple influencers}
\label{sec:3}

In spite of the great value of estimating individual nodes' influence with centralities, in a realistic situation, it is more relevant to understand spreading processes initiated by several spreaders. In applications such as viral marketing, it is expected that the spreaders can be coordinated in an optimal manner so that the final collective influence will be maximized. Although it sounds similar to the problem of locating single influencers, the collective influence maximization is in fact a fundamentally different and more difficult problem. In the seminal work of Kempe {\it et al.} \cite{kempe2003maximizing}, the influence maximization problems in both Independent Cascade Model (ICM) and Linear Threshold Model (LTM) were mapped to the NP-complete Vertex Cover problem. This implies, the influence maximization problem cannot be solved exactly within a polynomial time, leaving us the only choice of heuristic approach.

A straightforward idea to find multiple influencers is to select the top-ranked spreaders as individual seeds using centrality measures. However, this approach neglects the interactions and collective effect among spreaders. As demonstrated in SIR simulations, the selected spreaders have significant overlap in their influenced population \cite{kitsak2010identification}. Therefore, the set of influencers identified with centrality metrics are usually far from optimal. To solve this conundrum, it needs to be treated from a collective point of view \cite{morone2015influence}.

\subsection{Optimal percolation}
\label{sec:3.1}

We start our discussion from the percolation model point of view. As a well-studied dynamical process, percolation was shown to be closely related to spreading and immunization \cite{newman2002spread,pastor2002immunization,callaway2000network}. Percolation is a classical physical process in which nodes or links are randomly removed from a graph \cite{stauffer1994introduction}. The critical quantity that is of particular interest is the fraction of nodes or links whose removal will collapse the giant component. It is well known that the size of giant component decreases continuously to zero as the number of removed nodes or links increases. In the pioneering works of Newman \cite{newman2002spread,newman2001random}, the class of SIR models were mapped to the percolation process for which the critical point of the continuous transition could be solved exactly.

In contrast to the studies focused on random removal, the problem of optimal percolation aims to find the minimal set of nodes that could guarantee the global connectivity of the network, or equivalently, dismantle the network if removed. Morone and Makse showed that, mathematically, the optimization of spreading process following {\it exactly} the Linear Threshold Model with threshold $k-1$ ($k$ is the degree of each node) can be mapped to the optimal percolation problem \cite{morone2015influence}. For this specific spreading model, finding the minimum number of seeds so that the information percolates the entire network is essentially equivalent to locating the optimal set of nodes in the optimal percolation problem. Similarly, the optimal immunization problem, dual of optimal spreading, can also be mapped to optimal percolation \cite{morone2015influence}. The relation between the cohesion of a network and influence spreading indicates that the most influential spreaders are the nodes that maintain the integrity of the network.

The collective influence theory for optimal percolation is developed based on the message passing equations of the percolation process. For a network with $N$ nodes and $M$ edges, suppose $\mathbf{n}=(n_i,\cdots,n_N)$ indicates whether node $i$ is removed ($n_i=0$) or left ($n_i=1$) in the network. The total fraction of removed nodes is therefore $q=1-\sum_{i=1}^Nn_i/N$. For a directed link from $i$ to $j$ ($i\to j$), let $\nu_{i\to j}$ denote the probability of node $i$ belonging to the giant component $G$ in the absence of node $j$. The evolution of $\nu_{i\to j}$ satisfies the following self-consistent equation:
\begin{equation}\label{percolationmp1}
\nu_{i\to j}=n_i\left [1-\prod_{k\in\partial i \setminus j}(1-\nu_{k\to i})\right ],
\end{equation}
where $\partial i\setminus j$ denotes the nearest neighbors of $i$ excluding $j$. The final probability $\nu_i$ of node $i$ belonging to the giant component is then determined by $\nu_{k\to i}$ ($k\in\partial i$) through
\begin{equation}\label{percolationmp2}
\nu_{i}=n_i\left [1-\prod_{k\in\partial i}(1-\nu_{k\to i})\right ].
\end{equation}
The fraction of nodes in the giant component is then given by $G(q)=\sum_{i=1}^N\nu_i/N$.

\begin{figure}[t]\centering
\includegraphics[width=1\columnwidth]{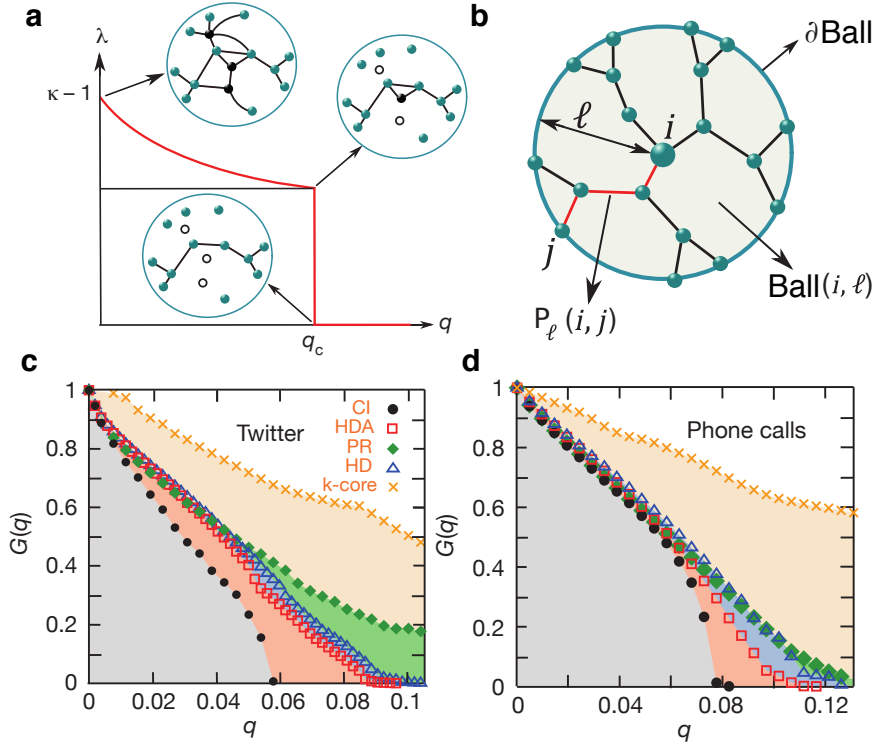}
\caption{{\bf a}, For $q\geq q_c$, the global minimum of the largest eigenvalue $\lambda$ of the NB matrix over $\mathbf{n}$ is 0. In this case, $G=0$ is stable, although there exist non-optimal configurations with $\lambda>1$ for which $G>0$. For $q<q_c$, the minimum of the largest eigenvalue is always $\lambda>1$. Therefore the solution $G=0$ is unstable and $G>0$. At the optimal percolation transition, the minimum is at $\mathbf{n}^*$ such that $\lambda(\mathbf{n}^*,q_c)=1$. At $q=0$, $\lambda=\kappa-1$ where $\kappa=\langle k^2\rangle/\langle k\rangle.$ At $\lambda=1$, the giant component is reduced to a tree plus one single loop. This loop is destroyed at the transition $q_c$, and $\lambda$ abruptly falls to 0. {\bf b}, $\text{Ball}(i,\ell)$ of radius $\ell$ around node $i$ is shown. $\partial\text{Ball}$ is the set of nodes on the boundary. The highlighted route is the shortest path from $i$ to $j$. {\bf c-d}, Giant component $G(q)$ of Twitter ($N=469,014$) and Mobile phone network in Mexico ($N=1.4\times 10^7$) computed using CI, high degree adaptive (HDA), PageRank (PR), high degree (HD) and k-core strategies. Figure is adapted from Morone {\it et al.} \cite{morone2015influence}.}
\label{fig:2}
\end{figure}

For the continuous phase transition in percolation process, the stability of the zero solution $G=0$ is determined by the largest eigenvalue $\lambda(\mathbf{n};q)$ of the coupling matrix $\mathcal{M}$ for the linearized Eq. (\ref{percolationmp1}) evaluated at $\{\nu_{i\to j}=0\}$ (see Fig. \ref{fig:2}a). Concretely, $\mathcal{M}$ is defined on the $2M\times 2M$ directed links as $\mathcal{M}_{k\to\ell,i\to j}\equiv\frac{\partial\nu_{i\to j}}{\partial\nu_{k\to\ell}}|_{\{\nu_{i\to j}=0\}}$. A simple calculation reveals that for locally-tree like random networks, $\mathcal{M}$ is given in terms of the Non-Backtracking (NB) matrix $\mathcal{B}$ \cite{hashimoto1989zeta} via $\mathcal{M}_{k\to\ell,i\to j}=n_i\mathcal{B}_{k\to\ell,i\to j}$ in which $\mathcal{B}_{k\to\ell,i\to j}=1$ if $\ell=i \text{ and } j\neq k$, and $0$ otherwise.

To guarantee the stability of the solution $\{\nu_{i\to j}=0\}$, it is required $\lambda(\mathbf{n};q)\leq1$. The optimal influence problem for a given $q$ can be rephrased as finding the optimal configuration $\mathbf{n}$ that minimizes the largest eigenvalue $\lambda(\mathbf{n}; q)$. As $q$ approaches the optimal threshold $q_c$, there exist a decreasing number of configurations that satisfy $\lambda(\mathbf{n};q)\leq1$. At $q_c$, only one configuration $\mathbf{n}^*$ exists such that $\lambda(\mathbf{n}^*;q_c)=1$, and all other configurations will give $\lambda(\mathbf{n};q)>1$. The optimal configuration of $Nq_c$ influencers $\mathbf{n}^*$ is therefore obtained when the {\it minimum} of the largest eigenvalue satisfies $\lambda(\mathbf{n}^*;q_c)=1$. In practice, the largest eigenvalue can be calculated by the power method (we leave out $q$ in $\lambda(\mathbf{n}; q)$):
\begin{equation}\label{powermethod}
\lambda(\mathbf{n})=\lim_{\ell\to\infty}\left[\frac{|\mathbf{w}_\ell(\mathbf{n})|}{|\mathbf{w}_0|}\right]^{1/\ell}.
\end{equation}
Here $|\mathbf{w}_\ell(\mathbf{n})|$ is the $\ell$ iterations of $\mathcal{M}$ on initial vector $\mathbf{w}_0$: $|\mathbf{w}_\ell(\mathbf{n})|=|\mathcal{M}^\ell\mathbf{w}_0|$. To find the best configuration of $\mathbf{n}$, we need to minimize the cost function $|\mathbf{w}_\ell(\mathbf{n})|$ for a finite $\ell$. Through a proper simplification, we have an approximation of $|\mathbf{w}_\ell(\mathbf{n})|^2$ of order $1/N$ as
\begin{equation}\label{wapprox}
|\mathbf{w}_\ell(\mathbf{n})|^2=\sum_{i=1}^N(k_i-1)\sum_{j\in\partial\text{Ball}(i,2\ell-1)}\left(\prod_{k\in\mathcal{P}_{2\ell-1}(i,j)}n_k\right)(k_j-1),
\end{equation}
in which $\partial\text{Ball}(i,\ell)$ is the frontier of the ball of radius $\ell$ in terms of shortest path centered around node $i$, $\mathcal{P}_\ell(i,j)$ is the shortest path of length $\ell$ connecting $i$ and $j$, and $k_i$ is the degree of node $i$. See an example in Fig. \ref{fig:2}b.

Based on the form of Eq. (\ref{wapprox}), an energy function for each configuration $\mathbf{n}$ can be defined as follows:
\begin{equation}\label{energy}
E_\ell(\mathbf{n})=\sum_{i=1}^N(k_i-1)\sum_{j\in\partial\text{Ball}(i,\ell)}\left(\prod_{k\in\mathcal{P}_{\ell}(i,j)}n_k\right)(k_j-1),
\end{equation}
where $E_\ell(\mathbf{n})=|\mathbf{w}_{(\ell+1)/2}|^2$ for $\ell$ odd and $E_\ell(\mathbf{n})=\langle \mathbf{w}_{\ell/2}|\mathcal{M}|\mathbf{w}_{\ell/2}\rangle$ for $\ell$ even. For $\ell=1$, $E_\ell(\mathbf{n})$ is exactly the energy function of an Ising model which can be optimized using the cavity method \cite{mezard2003cavity}. For $\ell\geq2$, it becomes a hard optimization problem involving many-body interactions. To develop a scalable algorithm for big-data analysis, an adaptive method is proposed, which is essentially a greedy algorithm for minimizing the largest eigenvalue of the stability matrix $\mathcal{M}$ for a given $\ell$ in the form of Eq. (\ref{wapprox}). In fact, Eq. (\ref{energy}) can be rewritten as the sum of collective influence from single nodes:
\begin{equation}\label{energyCI}
E_\ell(\mathbf{n})=\sum_{i=1}^N\text{CI}(i),
\end{equation}
in which the collective influence (CI) of node $i$ at length $\ell$ is defined as:
\begin{equation}\label{CI}
\text{CI}_\ell(i)=(k_i-1)\sum_{j\in\partial\text{Ball}(i,\ell)}(k_j-1).
\end{equation}
The main idea behind the CI algorithm is to remove the nodes that can cause largest decrease of energy function in Eq. (\ref{wapprox}). In each iteration of CI algorithm, the node with largest CI value is deleted, after which the CI values for remaining nodes are recalculated. The adaptive removal continues until the giant component is fragmented, i.e. $G(q)=0$. Notice that the procedure minimizes $q_c$ but does not guarantee the minimization of $G$ in the percolation phase $G>0$. If we want to optimize the configuration for $G(q)>0$, a reinsertion procedure is applied from the configuration at $G(q)=0$. In practice, if we use a heap structure to find the node with the largest CI and only update the nodes inside the $(\ell+1)$-radius ball around the removed node, the computational complexity of CI algorithm can achieve $N\log(N)$ \cite{morone2016collective}. As a result, the CI algorithm is scalable for massively large-scale networks in modern social network analysis. For a Twitter network with $469,013$ users (Fig. \ref{fig:2}c) and a social network of $1.4\times 10^7$ mobile phone users in Mexico (Fig. \ref{fig:2}d), CI algorithm finds a smaller set of influencers than simple scalable heuristics including high degree adaptive (HDA), PageRank (PR), high degree (HD), and k-core \cite{morone2015influence}. To apply CI algorithm to real-time influencer ranking, a Twitter search engine was developed at \url{http://www.kcore-analytics.com}. Notice that, for $\ell=0$, CI algorithm degenerates to high-degree ranking. So degree can be interpreted as the zero-oder approximation of CI in Eq. (\ref{CI}).

To guarantee the scalability of the algorithm, CI essentially takes an adaptive greedy approach. The performance of CI algorithm can be further improved by a simple extension of CI using the message passing framework for $\ell\to\infty$ - the CI propagation algorithm ($\text{CI}_\text{P}$) \cite{morone2016collective}. Remarkably, the CI propagation algorithm can reproduce the exact analytical threshold of optimal percolation for cubic random regular graphs \cite{bau2002decycling}. Another belief-propagation variant of CI algorithm based on optimal immunization ($\text{CI}_\text{BP}$) also has similar performance of $\text{CI}_\text{P}$ \cite{morone2016collective}. However, the improvement over CI algorithm is at the price of higher computational complexity $O(N^2\log(N))$, which makes both $\text{CI}_\text{P}$ and $\text{CI}_\text{BP}$ unscalable. 

Recent studies have shown that the optimal percolation problem is closely related to the optimal decycling problem, or minimum feedback vertex set (FVS) problem \cite{karp1972reducibility}. Using belief-propagation (BP) algorithms, the optimal percolation problem was solved in recent works \cite{mugisha2016identifying,braunstein2016network}. The result of BP algorithms was found better than CI algorithm. Another approach to the optimal destruction of networks makes use of the explosive percolation theory \cite{clusella2016immunization}.

\subsection{Independent Cascade Model}
\label{sec:3.2}

The percolation process is deterministic on a given network with a given seed set. An important class of spreading model with stochasticity is the independent cascade model (ICM) \cite{kleinberg2007cascading}. In these models, a node is infected or activated by its neighbors with a predefined probability independently. Frequently used independent cascade models include susceptible-infected (SI) model, susceptible-infected-susceptible (SIS) model and susceptible-infected-removed (SIR) model. These models are widely adopted in modeling infectious disease outbreaks and information spreading in social networks \cite{hethcote2000mathematics,kitsak2010identification,tang2015identification,pei2015detecting,yan2014dynamical,yan2015global}. Therefore, it is of particular interest in relevant applications.

In the pioneering work of Kempe {\it et al.} \cite{kempe2003maximizing}, influence maximization was first formalized as a discrete optimization problem: For a given spreading process on a network and an integer $k$, how to find the optimal set of $k$ seeds that could generate the largest influence. For a large class of ICM and LTM, the influence maximization problem can be well approximated by a simple greedy strategy, with a provable approximation guarantee \cite{kempe2003maximizing}. In the basic greedy algorithm, the seed set is obtained by repeatedly selecting the node that provides the largest marginal increase of influence at each time step. The performance guarantee is built on the submodular property of the influence function $\sigma(S)$ \cite{nemhauser1978analysis}, which is defined as the expected number of active nodes if the initial seed set is $S$. The influence function $\sigma(\cdot)$ is submodular if the incremental influence of selecting a node $u$ into a seed set $S$ is no smaller than the incremental influence of selecting the same node into a larger set $V$ containing $S$. That is, $\sigma(S\cup\{u\})-\sigma(S)\geq\sigma(V\cup\{u\})-\sigma(V)$ for all nodes $u$ and any sets $S\subseteq V$. Leveraging on the result of submodular function \cite{nemhauser1978analysis}, the greedy algorithm is guaranteed to approximate the true optimal influence within a factor of $1-1/e\approx 63\%$, i.e., $\sigma(S)\geq(1-1/e)\sigma(S^*)$, where $S$ is the seed set obtained by the greedy algorithm and $S^*$ is the true optimal seed set. Although the basic greedy algorithm is simple to implement and performance-guaranteed, it requires massive Monte Carlo simulations to estimate the marginal gain of each candidate node. Several works were proposed to improve the efficiency of greedy algorithm \cite{leskovec2007cost,goyal2011celf++,chen2009efficient,chen2010scalable}.

While performance guaranteed, from an optimization point of view, the greedy algorithm may be stuck into local optimum. This drawback can be solved by a more sophisticated message passing approach. Altarelli {\it et al.} developed the message passing algorithms (both belief-propagation (BP) and max-sum (MS)) for the problem of optimal immunization for SIR and SIS model \cite{altarelli2014containing}, which can be applied to general ICMs. From another point of view, the independent cascade model can be naturally mapped to a bond percolation. Hu {\it et al.} found that in a series of real-world networks, most SIR spreading would be restrained to a local area while global-scale spreading rarely occurs \cite{hu2015optimizing}. Using the bond percolation theory, a characteristic local length termed influence radius was revealed. They argue that the global spreading optimization problem in fact can be solved locally, with the knowledge of the local environment within the influence radius.

\subsection{Linear Threshold Model}
\label{sec:3.3}

Compared with independent cascade model, linear threshold model is more complex in the sense that a node's state is collectively determined by its neighbors' state. In a typical instance of LTM, each node $v$ is assigned with a threshold value $\theta_v$ and each link $(u,v)$ is assigned with a weight $w(u,v)$. During the cascade, a node is activated only if the sum of weights of its activated neighbors reaches the threshold value, i.e. $\sum_{u\in\partial v}w(u,v)\geq\theta_v$. In the case where the weights and thresholds are drawn uniformly from the interval $[0,1]$, LTM was proven to be submodular \cite{kempe2003maximizing}. Therefore, the influence maximization in this class of LTM can be well approximated by the greedy strategy, as we introduced in above section. However, even with the lazy forward update \cite{leskovec2007cost}, the algorithm is still unscalable for large networks. Chen {\it et al.} found a way to approximate the influence of a node in a local subgraph \cite{chen2010scalable}, and developed a scalable greedy algorithm. Goyal {\it et al.} \cite{goyal2011simpath} further improved this algorithm by considering more choices of paths.

The above greedy approach and its variants are applicable to LTM with submodular property. However, for the general class of LTM with fixed weight and threshold, it is not guaranteed to be submodular \cite{kempe2003maximizing}. An important class of LTM that may not be submodular is defined as follows: A node $i$ is activated only after a certain number $m_i$ of its neighbors are activated. The choice of different threshold $m_i$ can generate two qualitatively different cascade regimes with continuous and discontinuous phase transitions. For instance, in the special case of $m_i=k_i-1$ ($k_i$ is the degree of node $i$), a continuous phase transition of influence occurs as the seed set grows \cite{morone2015influence}. However, there also exist a wide class of LTM exhibiting a first-order, or discontinuous phase transition. In the case that seeds are selected randomly, the transition between these two regimes is explored in detail in the context of bootstrap percolation \cite{baxter2010bootstrap,goltsev2006k} and a simple cascade model \cite{watts2002simple}. But these results are based on the typical dynamical properties starting from random initial conditions. For influence maximization with a special initial condition, the dynamical behavior should be deviated from the average ones. Altarelli {\it et al.} proposed a BP algorithm that could estimate statistical properties of nontypical trajectories and found the initial conditions that lead to cascading with desired properties \cite{altarelli2013large}. To obtain the exact set of seeds, MS equations were derived by setting the inverse temperature $\beta\to\infty$ in the energy function \cite{altarelli2013optimizing}. Extending the work under the assumption of replica symmetry, the theoretical limit of the minimal contagious set (the minimal seed set that can activate the entire graph) in random regular graphs is obtained using the cavity method with the effect of replica symmetry breaking \cite{guggiola2015minimal}.

In big-data analysis, an efficient and scalable algorithm designed for general LTM is needed. Starting from the message passing equations of LTM, generalized from Eq. (\ref{percolationmp1}) of percolation, a scalable algorithm named collective influence for threshold model (CI-TM) can be developed \cite{pei2016efficient}. By iteratively solving the linearized message passing equations, the cascading process can be decomposed to separate components, each of which corresponds to the contribution made by a single seed. Interestingly, it is found the contribution of a seed is determined by the subcritical paths along which cascade propagates. In order to design a scalable algorithm, the node with the largest number of subcritical paths is recursively selected into the seed set. After each selection, the selected node and the subcritical paths attached to it are removed, and the status of the remaining nodes is recalculated. Making use of the heap structure, CI-TM algorithm can achieve the complexity of $O(N\log N)$. On one hand, computing $\text{CI-TM}_\ell$ value for a given length $\ell$ is equivalent to iteratively visiting subcritical neighbors of each node layer by layer within $\ell$ radius. Because of the finite search radius, computing $\text{CI-TM}_\ell$ for each node takes $O(1)$ time. Initially, we have to calculate $\text{CI-TM}_\ell$ for all nodes. However, during later adaptive calculation, there is no need to update $\text{CI-TM}_\ell$ for all nodes. We only have to recalculate for nodes within $\ell+1$ steps from the removed vertices, which scales as $O(1)$ compared to the network size as $N\to \infty$ as shown in  \cite{morone2016collective}. On the other hand, selecting the node with maximal CI-TM can be realized by making use of the data structure of heap that takes $O(\log N)$ time \cite{morone2016collective}. Therefore, the overall complexity of ranking $N$ nodes is $O(N\log N)$ even when we remove the top CI-TM nodes one by one. In both homogeneous and scale-free random networks, CI-TM achieves larger collective influence given the same number of seeds compared with other scalable approaches. This provides a practical method that can be applied to massively large-scale networks.

\section{Applications of influencer identification}
\label{sec:4}

The problem of influencer identification is ubiquitous in a wide class of applications. So far, the theory of influencer identification has been applied to a number of important problems. In this section, we will introduce the application of influencer identification in three different areas: information diffusion, brain networks, and socioeconomic systems.

\subsection{Information diffusion in social networks}
\label{sec:4.1}

The most direct application of influencer identification is to maximize the information diffusion in social networks. In recent years, a huge number of research works have been performed aiming to relate users' spreading power to their locations, or personal features \cite{pei2014searching,teng2016collective,min2015finding}. These works, mainly focusing on various types of online social networks including email communication \cite{liben2008tracing}, Facebook \cite{viswanath2009evolution,mislove2007measurement}, Twitter \cite{cha2010measuring,bakshy2011everyone,kwak2010twitter}, and blogs sharing communities \cite{backstrom2006group,ramos2014does}, enrich our understanding of information diffusion in social networks.

A great challenge of developing effective predictors of influencers comes from the validation. In most of the previous works, the validation of proposed measures depends on modeling of information spreading in a given network. This approach, however, has led to several contradictory results on the best predictor of influence depending on the particular models \cite{kitsak2010identification,borge2012absence}. These models are built on simplified assumptions on human behavior \cite{hu2014conditions} that neglect some of the most important features in real information diffusion \cite{gallos2008scaling}, such as activity frequency \cite{rybski2012communication,muchnik2013origins}, behavior pattern \cite{pei2015exploring,li2014rumor,teng2014individual}, etc. Therefore, it is required to validate the various proposed predictors using empirical diffusion records in real-world social media.

\begin{figure}[t]\centering
\includegraphics[width=1\columnwidth]{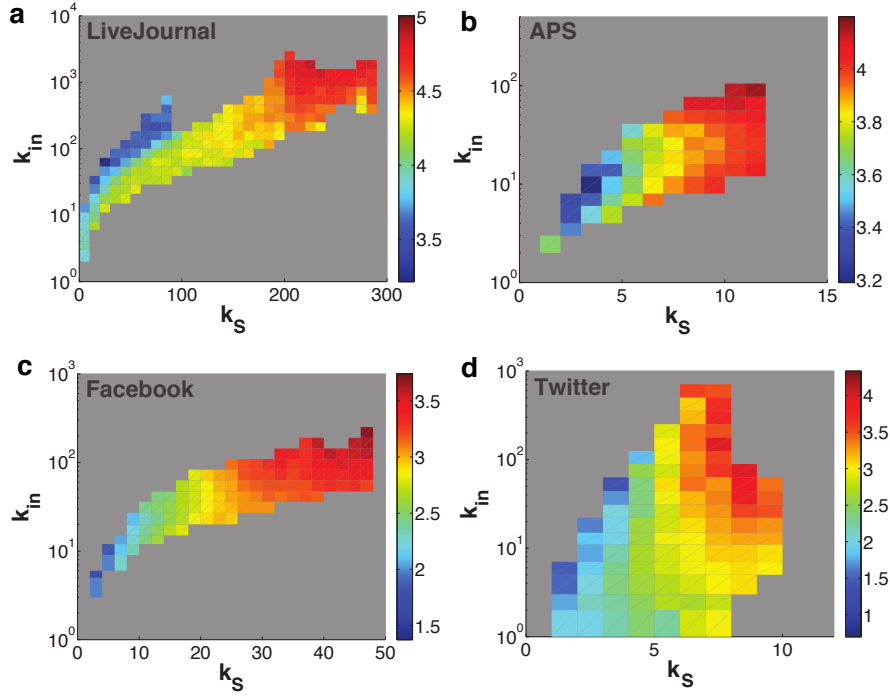}
\caption{K-core predicts the average influence of spreading more reliably than in-degree. Logarithmic values (base 10) of the average size of influence region $M(k_S,k_{in})$ when spreading originates from nodes with $(k_S,k_{in})$ for LiveJournal ({\bf a}), APS journals ({\bf b}), Facebook ({\bf c}) and Twitter ({\bf d}) are shown. Figure is adapted from Pei {\it et al.} \cite{pei2014searching}.}
\label{fig:3}
\end{figure}

We first compare the performance of different predictors for single influencers \cite{pei2014searching}. Realistic information diffusion instances as well as the underlying social networks are collected in four dissimilar social platforms: a blog-sharing community LiveJournal, scientific journals of American Physical Society, an online social network Facebook, and microblog service Twitter. To determine the real influence of each node, a directed diffusion graph is first constructed for each system by combining all directed diffusion links together. Then starting from a source node $i$, the total influence $M_i$ of node $i$ is computed by tracking the diffusion links layer by layer in a breadth-first-search (BFS) fashion. Once we get the realistic influence, it is convenient to compare the performance of different predictors, including degree, k-core, and PageRank. Specifically, we can calculate the average influence $M(k_S, k_{in})$ for nodes with a given combination of k-core value $k_S$ and in-degree $k_{in}$: $M(k_S, k_{in})=\sum_{i\in\Upsilon(k_S,k_{in})}M_i/N(k_S,k_{in})$, where $\Upsilon(k_S, k_{in})$ is the collection of users in the $(k_S, k_{in})$ bin, and $N(k_S, k_{in})$ is the size of this collection. In all the systems, it is consistently observed that nodes with fixed degree can have either large or small influence, while nodes located in the same k-core have similar influence (see Fig. \ref{fig:3}). Thus the influence of nodes is more related to their global location in the network, indicated by their k-core values. The same conclusion is also obtained in the comparison with PageRank. K-core does not only predict the average influence better, but also recognize influencers more accurately. Although k-core is effective, it is too coarse to distinguish different nodes within same shells. In some cases, there may be millions of nodes in one shell.

\begin{figure}[t]\centering
\includegraphics[width=1\columnwidth]{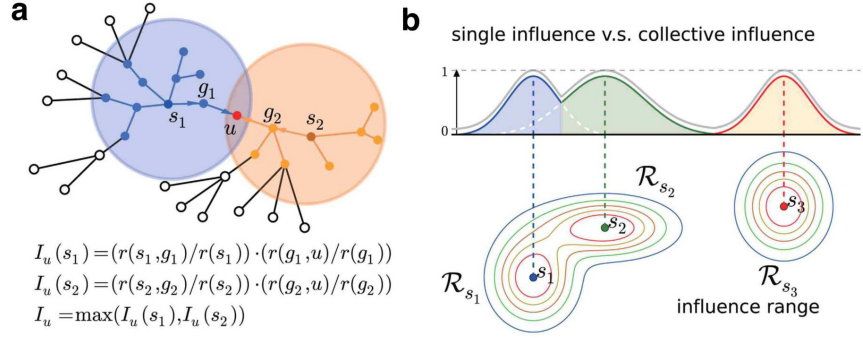}
\caption{ {\bf a}, Calculation of influence strength to node $u$. Suppose the maximum spreading layer is set as $L=2$ for two distinct seeds $s_1$ and $s_2$. The collective influence enforcing to $u$ is selected as the largest value of the strength $I_u(s_1)$ and $I_u(s_2)$. {\bf b}, An illustration of single influence and collective influence. The three circle-like areas represent influence range $R_{s_1}$, $R_{s_2}$ and $R_{s_3}$ for different spreaders $s_1$, $s_2$ and $s_3$. The contour lines show the levels of influence strength. The collective influence (grey curve) is obtained by combining single influence strengths of all spreaders. Figure is adapted from Teng {\it et al.} \cite{teng2016collective}.}
\label{fig:4}
\end{figure}

We further investigate the identification of multiple influencers \cite{teng2016collective}. Again, we use the realistic diffusion instances in the above four platforms. However, the empirical data cannot be directly mapped to ideal multi-source spreading. Such ideal multi-source spreading instances in which spreaders send out the same piece of message at the same time rarely exist in reality. Even though we can find such instances, the initial spreaders are hardly the same as the set of nodes selected by CI or other heuristic strategies. To circumvent this difficulty, we can construct virtual multi-source spreading processes by leveraging the behavior patterns of users extracted from the data. Suppose $n$ spreaders $S=\{s_i|i=1,2,\cdots,n, n=qN\}$ are activated at the beginning of the virtual process. The influence strength $I_{g_1}(s)$ from seed $s$ to its neighbor $g_1$ depends on the tendency of $g_1$ to receive information from $s$. Assume during the observation time, $s$ has sent out $r(s)$ pieces of messages and $g_1$ has accepted $r(s,g_1)$ of them. Then the influence strength can be approximated by $I_{g_1}(s)=r(s,g_1)/r(s)$. In subsequent spreading, $g_1$ may affect its neighbor $g_2\neq s$ in the same manner. Following the spreading paths, we can acquire the influence strength $s$ enforcing on its $\ell$-step neighbor $g_\ell$: $I_{g_\ell}(s)=\prod_{k=1}^\ell r(g_{k-1},g_k)/r(g_{k-1})$, where $g_0=s$. The collective influence $I_u$ for node $u$ imposed by the seed set $S$ is therefore $I_u=\max_{i=1}^n I_u(s_i)$. See Fig. \ref{fig:4} for an example. Finally, summing up all the $N$ nodes in the network, the collective influence of the spreaders imposed on the entire system is $Q(q)=\sum_{u=1}^N I_u/N$. Based on this virtual spreading process, we can evaluate the collective influence of the spreaders selected by different methods. In particular, we compare the influencers selected by collective influence algorithm (CI), adaptive high degree (HDA), high degree (HD), PageRank (PR), and k-core. In all the systems, CI consistently outperforms other ranking methods.

\subsection{Collective influence in brain networks}
\label{sec:4.2}

The human brain is a robust modular system interconnected as a Network of Networks (NoN) \cite{bullmore2009complex,reis2014avoiding,gallos2012small}. How this robustness emerges in a modular structure is an important question in many disciplines. Previous interdependent NoN models inspired by power grid are extremely fragile \cite{buldyrev2010catastrophic}, thus cannot explain the observed robustness in brain networks. To reveal the mechanism beneath this robustness, a NoN model is proposed which can afford inter-link functionality and remain robust at the same time \cite{morone2017model,roth2017emergence}. 

In NoN system, the links are classified into two types: inter-modular links that represent the mutual dependencies between modules and intra-modular links that do not involve in the inter-modular dependencies. Denote $\mathcal{S}(i)$ and $\mathcal{F}(i)$ as the set of nodes connected to node $i$ via intra-modular and inter-modular links, respectively. Suppose the variable state of node $i$ is $\sigma_i\in\{0,1\}$ (inactive or active), and the external input to node $i$ is $n_i\in\{0,1\}$ (no input or input). In the general activation model, the variable state is related to the input through $\sigma_i=n_i\left [1-\prod_{j\in\mathcal{F}(i)}(1-n_j)\right ]$. That is, the node $i$ is activated only if $i$ receives the input ($n_i=1$) and at least one of its neighbors connected with inter-modular links receives the input. In a robust brain network, for typical input configuration $\vec{n}=(n_1, \cdots, n_N)$, the giant (largest) component of the active nodes $G$ with $\sigma_i=1$ should be globally connected. Therefore, the robustness of the brain network can be characterized by the critical value $q_{rand}=1-\langle\vec{n}\rangle$ of zero inputs such that $G(q_{rand})=0$. Here the input configuration $\vec{n}$ is sampled from a flat distribution. Ideally, the robust NoN should have no disconnected phase, with a large value of $q_{rand}$ close to 1.

\begin{figure}[t]\centering
\includegraphics[width=1\columnwidth]{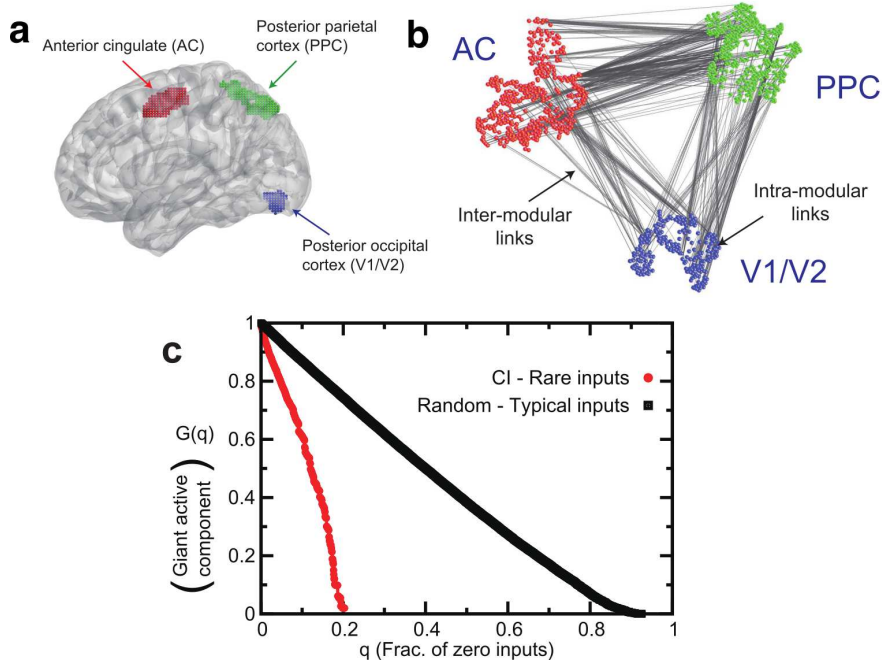}
\caption{ {\bf a}, Spatial location of the three main modules (AC, PPC, and V1/V2) in the 3NoN. {\bf b}, Topology of the 3NoN. Inter-links and intra-links are displayed. {\bf c}, Size of the largest active cluster $G(q)$ as a function $q$ of the nodes with $n_i=0$ following CI optimization (red curve, $\ell$=3) and random states (black curve, random percolation). Figure is adapted from Morone {\it et al.} \cite{morone2017model}.}
\label{fig:5}
\end{figure}

To explain both robustness and inter-link functionality of brain networks, a robust NoN (R-NoN) model is proposed \cite{morone2017model}. Define $\rho_{i\to j}\in\{0,1\}$ as the message running along an intra-modular link $i\to j$, $\varphi_{i\to j}\in\{0,1\}$ as the message running along an inter-modular link $i\to j$. The information flow follows the self-consistent equations 
\begin{eqnarray}
\rho_{i\to j} = \sigma_i\left [ 1-\prod_{k\in\mathcal{S}(i)\setminus j}(1-\rho_{k\to i})\prod_{\ell\in\mathcal{F}(i)}(1-\varphi_{\ell\to i})\right ],\label{RNoN1}\\
\varphi_{i\to j} = \sigma_i\left[ 1-\prod_{k\in\mathcal{S}(i)}(1-\rho_{k\to i})\prod_{\ell\in\mathcal{F}(i)\setminus j}(1-\varphi_{\ell\to i}) \right ]. \label{RNoN2}
\end{eqnarray}
The physical meaning of the above equations is easy to be interpreted. For instance, in Eq. (\ref{RNoN1}), a positive message $\rho_{i\to j}$ is transmitted from $i$ to $j$ in the same module if node $i$ is active $\sigma_i=1$ and if it receives at least one positive message from either a node $k$ in the same module $\rho_{k\to i}=1$ {\it or} a node $\ell$ in the other module $\varphi_{\ell\to i}=1$. Notice that, the logical OR is important since it is the basis of the robustness of R-NoN. The final probability of node $i$ belonging to the largest active component $G$ is
\begin{equation}\label{RNoN3}
\rho_i=\sigma_i\left[ 1-\prod_{k\in\mathcal{S}(i)}(1-\rho_{k\to i})\prod_{\ell\in\mathcal{F}(i)}(1-\varphi_{\ell\to i})\right].
\end{equation}
The size of $G$ is therefore $G=\langle\rho_i\rangle$. In the R-NoN model, the system is robust since a node can be active $\sigma_i=1$ even it does not belong to $G$. This prevents catastrophic cascading effects in the catastrophic C-NoN model inspired by power grid failure \cite{buldyrev2010catastrophic}. In the C-NoN model, a node remains functional only if it belongs to the giant component in {\it both} networks. This implies the status of a node in one network is interdependent on its status in the other network. The fundamental difference between C-NoN and R-NoN is that, in C-NoN model, the size of $G$ is computed through
\begin{equation}\label{CNoN}
\rho_i=\sigma_i\left[ 1-\prod_{k\in\mathcal{S}(i)}(1-\rho_{k\to i})\right ]\left [1-\prod_{\ell\in\mathcal{F}(i)}(1-\varphi_{\ell\to i})\right].
\end{equation}
So the logical OR in Eq. (\ref{RNoN3}) is replaced by the logical AND in C-NoN. This stricter condition makes the system extremely sensitive to small perturbations. In synthetic NoN made of ER and SF random graphs, it is found the percolation threshold $q_{rand}$ of R-NoN model is close to 1. On the contrary, the C-NoN model has threshold $q_{rand}$ close to 0. This indicates that the two models indeed capture two different phenomena.

After exploring the behavior of R-NoN model under typical inputs, it is required to study the response to rare events targeting the influencers in the brain networks. Rare inputs $\{n_i=0\}$ targeting influencers may interrupt the global communication in the brain, which have been conjectured be responsible for certain neurological disorders. Or conversely, activating the influencers would optimally broadcast information to the entire network. Therefore, it is important to predict the location of the most influential nodes involved in information processing in the brain. To find the minimal fraction of nodes $q_{infl}$ in the brain network whose removal would optimally fragment the giant component, the R-NoN model is mapped to the optimal percolation. The collective influence of nodes is calculated by minimizing the largest eigenvalue of the modified NB matrix. Particularly, the collective influence of node $i$ is given by
\begin{equation}\label{CIBrain}
\text{CI}_{\ell}(i)=z_i\sum_{j\in\partial\text{Ball}(i,\ell)}z_j+\sum_{j\in\mathcal{F}(i): k_j^{out}=1}z_j\sum_{m\in\partial\text{Ball}(j,\ell)}z_m,
\end{equation}
where $z_i\equiv k_i^{in}+k_i^{out}-1$. The first term is the node-centric contribution, which presents in the single network case of optimal percolation, while the second term is the node-eccentric contribution, which is a new feature of the brain NoN.

Applying the R-NoN model and collective influence theory to real brain networks, it is possible to obtain the collective influence map of brain NoN. The brain network is constructed from the functional magnetic resonance imaging (fMRI) data of the experiment of stimulus driven attention \cite{morone2017model,gallos2012small,gallos2007conundrum,min2016finding}. In the experiment, each subject performs a dual visual-auditory task when receiving a visual stimulus and an auditory pitch simultaneously. This experiment requires the deployment of high level control modules in the brain, thus captures the role of dependency inter-modular connections. In the obtained brain network (see Fig. \ref{fig:5}a-b), it is observed that the system is robust with large threshold $q_{rand}\approx 0.9$. While the minimal set of influencers only requires $q_{infl}\approx 0.2$ fraction of nodes (see Fig. \ref{fig:5}c). Using the CI-map of the brain network, it is confirmed that control is deployed from the higher level module (Anterior cingulate) towards certain strategic locations in the lower ones (posterior parietal cortex, posterior occipital cortex). Moreover, the coarse-grain of the NoN to top CI nodes can predict the strategic areas in the brain.

\subsection{Financial status in socioeconomic systems}
\label{sec:4.3}

It has long been recognized that the pattern of individuals' social connection in society can affect people's financial status \cite{granovetter1973strength}. However, how to quantify the relationship between the location of an individual in social network and his/her economic wellness remains an open question. Despite that the effect of network diversity on economic development has been tested in the community level \cite{eagle2010network}, inference of people's financial status from social network centralities or metrics in individual level is still needed. The difficulty of such investigation comes from the lack of empirical data containing both individual's financial information and pattern of social ties.

\begin{figure}[t]\centering
\includegraphics[width=1\columnwidth]{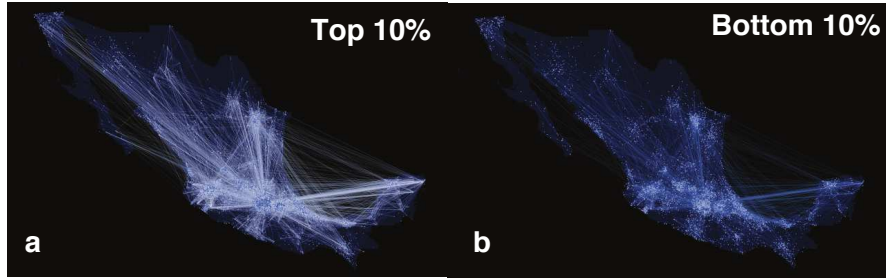}
\caption{ {\bf a-b}, Visualization of communication activity of population in the top $10\%$ and bottom $10\%$ total credit limit classes. Figure is adapted from Luo {\it et al.} \cite{luo2016inferring}.}
\label{fig:6}
\end{figure}

To find a reliable social network predictor of people's financial status, a massively large social network of the mobile and residential communication in Mexico containing $1.10\times 10^8$ users together with financial banking data are analyzed \cite{luo2016inferring}. With this dataset, it is possible to precisely cross-correlate the financial information of a person with his/her location in the communication network at the country level. Particularly, the financial status of individuals is reflected by their credit limit. In the analysis of the $5.02\times 10^5$ bank clients identified in the phone call network, the top $10\%$ and bottom $10\%$ individuals present completely different communication pattern (see Fig. \ref{fig:6}). Richer people maintain more active and diverse links, some connecting to remote locations and forming tightly linked ``rich clubs". 

To characterize the affluent people with network metrics, several centralities that are feasible for large-scale networks are compared, including degree, PageRank, k-core, and collective influence (CI). In the communication network, these four metrics are correlated. Therefore, they all show correlations with financial status when age is controlled. Among them, both k-core and CI capture the strong correlation with credit line with a $R^2$ value of 0.96 and 0.93, respectively. However, CI is more preferable since it satisfies both, a strong correlation and a high resolution. According to the definition of CI, top CI nodes are surrounded by hubs hierarchically. This is exactly the structure of ego-centric network of the top $1\%$ wealthy people.

The performance of predictions can be further enhanced by considering the factor of age. An age-network combined metric $\text{ANC}=\alpha\text{Age}+(1-\alpha)\text{CI}$ with $\alpha=0.5$ can achieve a correlation with $R^2=0.99$. Moreover, it is able to identify $70\%$ high credit individuals at the highest earner level. To validate the effectiveness, a real social marketing campaign was performed. Specifically, text messages inviting new credit card clients were sent to 656,944 people selected by their high CI values in the social network. Meanwhile, the same message was sent to a control group of 48,000 individuals selected randomly. The response rate, measured by the fraction of recipients who requested the product, is augmented by threefold in the top influencers identified by CI compared with the random control group.

The same analysis was also applied to individuals' diversity of links \cite{eagle2010network}. The diversity of an individual can be measured by the diversity ratio $\text{DR}=W_{out}/W_{in}$, i.e., the ratio of total communication events with people in other communities $W_{out}$ and within the same community $W_{in}$. The correlation between DR and CI is weak so they should reflect different aspects of network structure. In comparison with financial data, the age-diversity composite $\text{ADC}=\alpha\text{Age}+(1-\alpha)\text{DR}$ ($\alpha=0.5$) well correlates with people's financial status. These evidences indicate that both CI and DR are effective predictors of people's financial situation in an individual level. This finding has a great practical value in relevant applications, for instance, social marketing campaigns.

\section{Acknowledgement}
\label{sec:5}
We acknowledge funding from NIH-NIBIB 1R01EB022720, NIH-NCI U54CA137788 / U54CA132378 and NSF-IIS 1515022.

\bibliography{ref} 

\begin{thebibliography}{10}
\providecommand{\url}[1]{{#1}}
\providecommand{\urlprefix}{URL }
\expandafter\ifx\csname urlstyle\endcsname\relax
  \providecommand{\doi}[1]{DOI~\discretionary{}{}{}#1}\else
  \providecommand{\doi}{DOI~\discretionary{}{}{}\begingroup
  \urlstyle{rm}\Url}\fi

\bibitem{albert2000error}
Albert, R., Jeong, H., Barab{\'a}si, A.L.: Error and attack tolerance of
  complex networks.
\newblock Nature \textbf{406}(6794), 378--382 (2000)

\bibitem{altarelli2014containing}
Altarelli, F., Braunstein, A., Dall’Asta, L., Wakeling, J.R., Zecchina, R.:
  Containing epidemic outbreaks by message-passing techniques.
\newblock Phys. Rev. X \textbf{4}(2), 021024 (2014)

\bibitem{altarelli2013large}
Altarelli, F., Braunstein, A., Dall’Asta, L., Zecchina, R.: Large deviations
  of cascade processes on graphs.
\newblock Phys. Rev. E \textbf{87}(6), 062115 (2013)

\bibitem{altarelli2013optimizing}
Altarelli, F., Braunstein, A., Dall’Asta, L., Zecchina, R.: Optimizing spread
  dynamics on graphs by message passing.
\newblock J. Stat. Mech: Theory and Exp. \textbf{2013}(09), P09011 (2013)

\bibitem{backstrom2006group}
Backstrom, L., Huttenlocher, D., Kleinberg, J., Lan, X.: Group formation in
  large social networks: membership, growth, and evolution.
\newblock In: Proc.12th ACM SIGKDD Intl. Conf. on Knowledge Discovery and Data
  Mining, pp. 44--54. ACM (2006)

\bibitem{bakshy2011everyone}
Bakshy, E., Hofman, J.M., Mason, W.A., Watts, D.J.: Everyone's an influencer:
  quantifying influence on twitter.
\newblock In: Proc. 4th ACM Intl. Conf. on Web Search and Data Mining, pp.
  65--74. ACM (2011)

\bibitem{batagelj2003m}
Batagelj, V., Zaversnik, M.: An o (m) algorithm for cores decomposition of
  networks.
\newblock arXiv preprint cs/0310049  (2003)

\bibitem{bau2002decycling}
Bau, S., Wormald, N.C., Zhou, S.: Decycling numbers of random regular graphs.
\newblock Random Struct. Alg. \textbf{21}(3-4), 397--413 (2002)

\bibitem{baxter2010bootstrap}
Baxter, G.J., Dorogovtsev, S.N., Goltsev, A.V., Mendes, J.F.: Bootstrap
  percolation on complex networks.
\newblock Phys. Rev. E \textbf{82}(1), 011103 (2010)

\bibitem{bonacich1972factoring}
Bonacich, P.: Factoring and weighting approaches to status scores and clique
  identification.
\newblock J. Math. Socio. \textbf{2}(1), 113--120 (1972)

\bibitem{borge2012absence}
Borge-Holthoefer, J., Moreno, Y.: Absence of influential spreaders in rumor
  dynamics.
\newblock Phys. Rev. E \textbf{85}(2), 026116 (2012)

\bibitem{braunstein2016network}
Braunstein, A., Dall’Asta, L., Semerjian, G., Zdeborov{\'a}, L.: Network
  dismantling.
\newblock Proc. Natl. Acad. Sci. U.S.A. \textbf{113}(44), 12,368--12,373 (2016)

\bibitem{brin1998anatomy}
Brin, S., Page, L.: The anatomy of a large-scale hypertextual web search
  engine.
\newblock Computer Networks and ISDN System \textbf{30}(1), 107--117 (1998)

\bibitem{buldyrev2010catastrophic}
Buldyrev, S.V., Parshani, R., Paul, G., Stanley, H.E., Havlin, S.: Catastrophic
  cascade of failures in interdependent networks.
\newblock Nature \textbf{464}(7291), 1025--1028 (2010)

\bibitem{bullmore2009complex}
Bullmore, E., Sporns, O.: Complex brain networks: graph theoretical analysis of
  structural and functional systems.
\newblock Nat. Rev. Neurosci. \textbf{10}(3), 186--198 (2009)

\bibitem{callaway2000network}
Callaway, D.S., Newman, M.E., Strogatz, S.H., Watts, D.J.: Network robustness
  and fragility: Percolation on random graphs.
\newblock Phys. Rev. Lett. \textbf{85}(25), 5468 (2000)

\bibitem{centola2010spread}
Centola, D.: The spread of behavior in an online social network experiment.
\newblock Science \textbf{329}(5996), 1194--1197 (2010)

\bibitem{cha2010measuring}
Cha, M., Haddadi, H., Benevenuto, F., Gummadi, P.K.: Measuring user influence
  in twitter: The million follower fallacy.
\newblock Proc. 4th Intl. AAAI Conf. on Weblogs and Social Media
  \textbf{10}(10-17), 30 (2010)

\bibitem{chen2010scalable}
Chen, W., Wang, C., Wang, Y.: Scalable influence maximization for prevalent
  viral marketing in large-scale social networks.
\newblock In: Proc. 16th ACM SIGKDD Intl. Conf. on Knowledge Discovery and Data
  Mining, pp. 1029--1038. ACM (2010)

\bibitem{chen2009efficient}
Chen, W., Wang, Y., Yang, S.: Efficient influence maximization in social
  networks.
\newblock In: Proc. 15th ACM SIGKDD Intl. Conf. on Knowledge Discovery and Data
  Mining, pp. 199--208. ACM (2009)

\bibitem{clusella2016immunization}
Clusella, P., Grassberger, P., P{\'e}rez-Reche, F.J., Politi, A.: Immunization
  and targeted destruction of networks using explosive percolation.
\newblock Phys. Rev. Lett. \textbf{117}(20), 208301 (2016)

\bibitem{cohen2001breakdown}
Cohen, R., Erez, K., Ben-Avraham, D., Havlin, S.: Breakdown of the internet
  under intentional attack.
\newblock Phys. Rev. Lett. \textbf{86}(16), 3682 (2001)

\bibitem{eagle2010network}
Eagle, N., Macy, M., Claxton, R.: Network diversity and economic development.
\newblock Science \textbf{328}(5981), 1029--1031 (2010)

\bibitem{min2016finding}
del Ferraro, G., Moreno, A., Min, B., Morone, F., Perez-Ramirez, U.,
  Perez-Cervera, L., Parra, L., A, H., Canals, S., Makse, H.A.: Finding
  essential nodes for integration in the brain using network optimization
  theory  (2017)

\bibitem{freeman1978centrality}
Freeman, L.C.: Centrality in social networks conceptual clarification.
\newblock Soc. Netw. \textbf{1}(3), 215--239 (1978)

\bibitem{gallos2012small}
Gallos, L.K., Makse, H.A., Sigman, M.: A small world of weak ties provides
  optimal global integration of self-similar modules in functional brain
  networks.
\newblock Proc. Natl. Acad. Sci. U.S.A. \textbf{109}(8), 2825--2830 (2012)

\bibitem{gallos2007conundrum}
Gallos, L.K., Sigman, M., Makse, H.A.: The conundrum of functional brain
  networks: small-world efficiency or fractal modularity.
\newblock Front. Psychol. \textbf{3}, 123 (2007)

\bibitem{gallos2008scaling}
Gallos, L.K., Song, C., Makse, H.A.: Scaling of degree correlations and its
  influence on diffusion in scale-free networks.
\newblock Phys. Rev. Lett. \textbf{100}(24), 248,701 (2008)

\bibitem{goltsev2006k}
Goltsev, A.V., Dorogovtsev, S.N., Mendes, J.F.F.: k-core (bootstrap)
  percolation on complex networks: Critical phenomena and nonlocal effects.
\newblock Phys. Rev. E \textbf{73}(5), 056101 (2006)

\bibitem{goyal2011celf++}
Goyal, A., Lu, W., Lakshmanan, L.V.: Celf++: optimizing the greedy algorithm
  for influence maximization in social networks.
\newblock In: Proc. 20th Intl. Conf. World Wide Web, pp. 47--48. ACM (2011)

\bibitem{goyal2011simpath}
Goyal, A., Lu, W., Lakshmanan, L.V.: Simpath: An efficient algorithm for
  influence maximization under the linear threshold model.
\newblock In: Data Mining (ICDM), 2011 IEEE 11th Intl. Conf. on, pp. 211--220.
  IEEE (2011)

\bibitem{granovetter1973strength}
Granovetter, M.S.: The strength of weak ties.
\newblock Am. J. Sociol. \textbf{78}(6), 1360--1380 (1973)

\bibitem{guggiola2015minimal}
Guggiola, A., Semerjian, G.: Minimal contagious sets in random regular graphs.
\newblock J. Stat. Phys. \textbf{158}(2), 300--358 (2015)

\bibitem{hashimoto1989zeta}
Hashimoto, K.i.: Zeta functions of finite graphs and representations of p-adic
  groups.
\newblock Adv. Stud. Pure Math. \textbf{15}, 211--280 (1989)

\bibitem{hethcote2000mathematics}
Hethcote, H.W.: The mathematics of infectious diseases.
\newblock SIAM Rev. \textbf{42}(4), 599--653 (2000)

\bibitem{hu2014conditions}
Hu, Y., Havlin, S., Makse, H.A.: Conditions for viral influence spreading
  through multiplex correlated social networks.
\newblock Phys. Rev. X \textbf{4}(2), 021,031 (2014)

\bibitem{hu2015optimizing}
Hu, Y., Ji, S., Feng, L., Havlin, S., Jin, Y.: Optimizing locally the spread of
  influence in large scale online social networks.
\newblock arXiv preprint arXiv:1509.03484  (2015)

\bibitem{karp1972reducibility}
Karp, R.M.: Reducibility among combinatorial problems.
\newblock In: Complexity of computer computations, pp. 85--103. Springer (1972)

\bibitem{katz1953new}
Katz, L.: A new status index derived from sociometric analysis.
\newblock Psychometrika \textbf{18}(1), 39--43 (1953)

\bibitem{kempe2003maximizing}
Kempe, D., Kleinberg, J., Tardos, {\'E}.: Maximizing the spread of influence
  through a social network.
\newblock In: Proc. 9th ACM SIGKDD Intl. Conf. on Knowledge Discovery and Data
  Mining, pp. 137--146. ACM (2003)

\bibitem{kitsak2010identification}
Kitsak, M., Gallos, L.K., Havlin, S., Liljeros, F., Muchnik, L., Stanley, H.E.,
  Makse, H.A.: Identification of influential spreaders in complex networks.
\newblock Nat. Phys. \textbf{6}(11), 888--893 (2010)

\bibitem{kleinberg2007cascading}
Kleinberg, J.: Cascading behavior in networks: Algorithmic and economic issues.
\newblock Algorithmic game theory \textbf{24}, 613--632 (2007)

\bibitem{klemm2012measure}
Klemm, K., Serrano, M., Eguiluz, V.M., Miguel, M.S.: A measure of individual
  role in collective dynamics.
\newblock Sci. Rep. \textbf{2}, 292 (2012)

\bibitem{kwak2010twitter}
Kwak, H., Lee, C., Park, H., Moon, S.: What is twitter, a social network or a
  news media?
\newblock In: Proc.19th ACM Intl. Conf. on World Wide Web, pp. 591--600. ACM
  (2010)

\bibitem{lawyer2015understanding}
Lawyer, G.: Understanding the influence of all nodes in a network.
\newblock Sci. Rep. \textbf{5}, 8665 (2015)

\bibitem{leskovec2007dynamics}
Leskovec, J., Adamic, L.A., Huberman, B.A.: The dynamics of viral marketing.
\newblock ACM Trans. Web \textbf{1}(1), 5 (2007)

\bibitem{leskovec2007cost}
Leskovec, J., Krause, A., Guestrin, C., Faloutsos, C., VanBriesen, J., Glance,
  N.: Cost-effective outbreak detection in networks.
\newblock In: Proc. 13th ACM SIGKDD Intl. Conf. on Knowledge Discovery and Data
  Mining, pp. 420--429. ACM (2007)

\bibitem{li2014rumor}
Li, W., Tang, S., Pei, S., Yan, S., Jiang, S., Teng, X., Zheng, Z.: The rumor
  diffusion process with emerging independent spreaders in complex networks.
\newblock Physica A \textbf{397}, 121--128 (2014)

\bibitem{liben2008tracing}
Liben-Nowell, D., Kleinberg, J.: Tracing information flow on a global scale
  using internet chain-letter data.
\newblock Proc. Natl. Acad. Sci. U.S.A. \textbf{105}(12), 4633--4638 (2008)

\bibitem{liu2014core}
Liu, Y., Tang, M., Zhou, T., Do, Y.: Core-like groups result in invalidation of
  identifying super-spreader by k-shell decomposition.
\newblock Sci. Rep. \textbf{5}, 9602 (2015)

\bibitem{liu2015improving}
Liu, Y., Tang, M., Zhou, T., Do, Y.: Improving the accuracy of the k-shell
  method by removing redundant links-from a perspective of spreading dynamics.
\newblock Sci. Rep. \textbf{5}, 13172 (2015)

\bibitem{lu2016vital}
L{\"u}, L., Chen, D., Ren, X.L., Zhang, Q.M., Zhang, Y.C., Zhou, T.: Vital
  nodes identification in complex networks.
\newblock Phys. Rep. \textbf{650}, 1--63 (2016)

\bibitem{lu2011leaders}
L{\"u}, L., Zhang, Y.C., Yeung, C.H., Zhou, T.: Leaders in social networks, the
  delicious case.
\newblock PLoS ONE \textbf{6}(6), e21202 (2011)

\bibitem{lu2016h}
L{\"u}, L., Zhou, T., Zhang, Q.M., Stanley, H.E.: The h-index of a network node
  and its relation to degree and coreness.
\newblock Nat. Comm. \textbf{7}, 10168 (2016)

\bibitem{luo2016inferring}
Luo, S., Morone, F., Sarraute, C., Makse, H.A.: Inferring personal financial
  status from social network location.
\newblock Nat. Comm. \textbf{8}, 15227 (2017)

\bibitem{martin2014localization}
Martin, T., Zhang, X., Newman, M.: Localization and centrality in networks.
\newblock Phys. Rev. E \textbf{90}(5), 052808 (2014)

\bibitem{mezard2003cavity}
M{\'e}zard, M., Parisi, G.: The cavity method at zero temperature.
\newblock J. Stat. Phys. \textbf{111}(1), 1--34 (2003)

\bibitem{min2015finding}
Min, B., Liljeros, F., Makse, H.A.: Finding influential spreaders from human
  activity beyond network location.
\newblock PLoS ONE \textbf{10}(8), e0136831 (2015)

\bibitem{min2016searching}
Min, B., Morone, F., Makse, H.A.: Searching for influencers in big-data complex
  networks.
\newblock In: Diffusive Spreading in Nature, Technology and Society (Springer
  Verlag, Edited by A. Bunde, J. Caro, J. Karger, G. Vogl) (2016)

\bibitem{mislove2007measurement}
Mislove, A., Marcon, M., Gummadi, K.P., Druschel, P., Bhattacharjee, B.:
  Measurement and analysis of online social networks.
\newblock In: Proc. 7th ACM SIGCOMM Conf. on Internet Measurement, pp. 29--42.
  ACM (2007)

\bibitem{morone2015influence}
Morone, F., Makse, H.A.: Influence maximization in complex networks through
  optimal percolation.
\newblock Nature \textbf{524}, 65--68 (2015)

\bibitem{morone2016collective}
Morone, F., Min, B., Bo, L., Mari, R., Makse, H.A.: Collective influence
  algorithm to find influencers via optimal percolation in massively large
  social media.
\newblock Sci. Rep. \textbf{6}, 30062 (2016)

\bibitem{morone2017model}
Morone, F., Roth, K., Min, B., Stanley, H.E., Makse, H.A.: A model of brain
  activation predicts the neural collective influence map of the human brain.
\newblock Proc. Natl. Acad. Sci. U.S.A. \textbf{114}(15), 3849--3854 (2017)

\bibitem{muchnik2013origins}
Muchnik, L., Pei, S., Parra, L.C., Reis, S.D., Andrade~Jr, J.S., Havlin, S.,
  Makse, H.A.: Origins of power-law degree distribution in the heterogeneity of
  human activity in social networks.
\newblock Sci. Rep. \textbf{3}, 1783 (2013)

\bibitem{mugisha2016identifying}
Mugisha, S., Zhou, H.J.: Identifying optimal targets of network attack by
  belief propagation.
\newblock Phys. Rev. E \textbf{94}(1), 012305 (2016)

\bibitem{nemhauser1978analysis}
Nemhauser, G.L., Wolsey, L.A., Fisher, M.L.: An analysis of approximations for
  maximizing submodular set functions—i.
\newblock Math. Program. \textbf{14}(1), 265--294 (1978)

\bibitem{newman2002spread}
Newman, M.E.: Spread of epidemic disease on networks.
\newblock Phys. Rev. E \textbf{66}(1), 016128 (2002)

\bibitem{newman2001random}
Newman, M.E., Strogatz, S.H., Watts, D.J.: Random graphs with arbitrary degree
  distributions and their applications.
\newblock Phys. Rev. E \textbf{64}(2), 026118 (2001)

\bibitem{pastor2001epidemic}
Pastor-Satorras, R., Vespignani, A.: Epidemic spreading in scale-free networks.
\newblock Phys. Rev. Lett. \textbf{86}(14), 3200 (2001)

\bibitem{pastor2002immunization}
Pastor-Satorras, R., Vespignani, A.: Immunization of complex networks.
\newblock Phys. Rev. E \textbf{65}(3), 036104 (2002)

\bibitem{pei2013spreading}
Pei, S., Makse, H.A.: Spreading dynamics in complex networks.
\newblock J. Stat. Mech: Theory Exp. \textbf{2013}(12), P12002 (2013)

\bibitem{pei2014searching}
Pei, S., Muchnik, L., Andrade~Jr, J.S., Zheng, Z., Makse, H.A.: Searching for
  superspreaders of information in real-world social media.
\newblock Sci. Rep. \textbf{4}, 5547 (2014)

\bibitem{pei2015exploring}
Pei, S., Muchnik, L., Tang, S., Zheng, Z., Makse, H.A.: Exploring the complex
  pattern of information spreading in online blog communities.
\newblock PLoS ONE \textbf{10}(5), e0126894 (2015)

\bibitem{pei2015detecting}
Pei, S., Tang, S., Zheng, Z.: Detecting the influence of spreading in social
  networks with excitable sensor networks.
\newblock PLoS ONE \textbf{10}(5), e0124,848 (2015)

\bibitem{pei2016efficient}
Pei, S., Teng, X., Shaman, J., Morone, F., Makse, H.A.: Efficient collective
  influence maximization in threshold models of behavior cascading with
  first-order transitions.
\newblock Sci. Rep. \textbf{7}, 45240 (2017)

\bibitem{radicchi2016leveraging}
Radicchi, F., Castellano, C.: Leveraging percolation theory to single out
  influential spreaders in networks.
\newblock Phys. Rev. E \textbf{93}(6), 062314 (2016)

\bibitem{ramos2014does}
Ramos, M., Shao, J., Reis, S.D., Anteneodo, C., Andrade~Jr, J.S., Havlin, S.,
  Makse, H.A.: How does public opinion become extreme?
\newblock Sci. Rep. \textbf{5}, 10032 (2015)

\bibitem{reis2014avoiding}
Reis, S.D., Hu, Y., Babino, A., Andrade~Jr, J.S., Canals, S., Sigman, M.,
  Makse, H.A.: Avoiding catastrophic failure in correlated networks of
  networks.
\newblock Nat. Phys. \textbf{10}(10), 762--767 (2014)

\bibitem{restrepo2006characterizing}
Restrepo, J.G., Ott, E., Hunt, B.R.: Characterizing the dynamical importance of
  network nodes and links.
\newblock Phys. Rev. Lett. \textbf{97}(9), 094102 (2006)

\bibitem{richardson2002mining}
Richardson, M., Domingos, P.: Mining knowledge-sharing sites for viral
  marketing.
\newblock In: Proc. 8th ACM SIGKDD Intl. Conf. on Knowledge Discovery and Data
  Mining, pp. 61--70. ACM (2002)

\bibitem{rogers2010diffusion}
Rogers, E.M.: Diffusion of innovations.
\newblock Simon and Schuster (2010)

\bibitem{roth2017emergence}
Roth, K., Morone, F., Min, B., Makse, H.A.: Emergence of robustness in networks
  of networks.
\newblock Phys. Rev. E \textbf{95}(6), 062,308 (2017)

\bibitem{rybski2012communication}
Rybski, D., Buldyrev, S.V., Havlin, S., Liljeros, F., Makse, H.A.:
  Communication activity in a social network: relation between long-term
  correlations and inter-event clustering.
\newblock Sci. Rep. \textbf{2}, 560 (2012)

\bibitem{sabidussi1966centrality}
Sabidussi, G.: The centrality index of a graph.
\newblock Psychometrika \textbf{31}(4), 581--603 (1966)

\bibitem{seidman1983network}
Seidman, S.B.: Network structure and minimum degree.
\newblock Soc. Netw. \textbf{5}(3), 269--287 (1983)

\bibitem{stauffer1994introduction}
Stauffer, D., Aharony, A.: Introduction to percolation theory.
\newblock CRC press (1994)

\bibitem{tang2015identification}
Tang, S., Teng, X., Pei, S., Yan, S., Zheng, Z.: Identification of highly
  susceptible individuals in complex networks.
\newblock Physica A \textbf{432}, 363--372 (2015)

\bibitem{teng2016collective}
Teng, X., Pei, S., Morone, F., Makse, H.A.: Collective influence of multiple
  spreaders evaluated by tracing real information flow in large-scale social
  networks.
\newblock Sci. Rep. \textbf{6}, 36043 (2016)

\bibitem{teng2014individual}
Teng, X., Yan, S., Tang, S., Pei, S., Li, W., Zheng, Z.: Individual behavior
  and social wealth in the spatial public goods game.
\newblock Physica A \textbf{402}, 141--149 (2014)

\bibitem{viswanath2009evolution}
Viswanath, B., Mislove, A., Cha, M., Gummadi, K.P.: On the evolution of user
  interaction in facebook.
\newblock In: Proc. 2nd ACM Workshop on Online Social Networks, pp. 37--42. ACM
  (2009)

\bibitem{watts2002simple}
Watts, D.J.: A simple model of global cascades on random networks.
\newblock Proc. Natl. Acad. Sci. U.S.A. \textbf{99}(9), 5766--5771 (2002)

\bibitem{watts2007influentials}
Watts, D.J., Dodds, P.S.: Influentials, networks, and public opinion formation.
\newblock J. Cons. Res. \textbf{34}(4), 441--458 (2007)

\bibitem{yan2015global}
Yan, S., Tang, S., Fang, W., Pei, S., Zheng, Z.: Global and local targeted
  immunization in networks with community structure.
\newblock J. Stat. Mech: Theory Exp. \textbf{2015}(8), P08010 (2015)

\bibitem{yan2014dynamical}
Yan, S., Tang, S., Pei, S., Jiang, S., Zheng, Z.: Dynamical immunization
  strategy for seasonal epidemics.
\newblock Phys. Rev. E \textbf{90}(2), 022808 (2014)

\bibitem{zeng2013ranking}
Zeng, A., Zhang, C.J.: Ranking spreaders by decomposing complex networks.
\newblock Phys. Lett. A \textbf{377}(14), 1031--1035 (2013)

\end{thebibliography}
\bibliographystyle{spmpsci}

\end{document}